\documentclass[aps,reprint,superscriptaddress,nofootinbib,showpacs,amsmath,amssymb]{revtex4-1}
\usepackage{latexsym,graphicx,slashed,bm,dcolumn}

\newcommand{\mat}[4]{\left(\begin{array}{cc}{#1}&{#2}\\{#3}&{#4}
\end{array}\right)}

\newcommand*\xbar[1]{%
\kern0.5ex%
 \hbox{%
  \kern0.2ex%
     \vbox{%
     \hrule height 0.5pt 
     \kern0.25ex
     \hbox{%
       \kern-0.1em
       \ensuremath{#1}%
       \kern-0.1em
     }%
   }%
 }%
}
\begin{document}
\begin{titlepage}
\begin{flushright}
FTPI-MINN-15/29\\[2mm]
NSF-KITP-15-073~~
\end{flushright}

\vspace{2cm}

\begin{center}
{  \Large \bf  Neutron--Antineutron Oscillation as a Signal of CP Violation}
\end{center}

\vspace{1.0cm}

\begin{center}
{\large
Zurab Berezhiani$^{1,2}$ and Arkady Vainshtein$^{\,3,4,5}$}
\end {center}

\vspace{1mm}

\begin{center}
$^{1}${\it Dipartimento di Fisica, Universit\`a dell'Aquila, Via Vetoio, 67100 Coppito, L'Aquila, Italy}\\[1mm]
$^{2}${\it INFN, Laboratori Nazionali Gran Sasso, 67010 Assergi,  L'Aquila, Italy}\\[1mm]
$^{3}${\it  Department of Physics, University of Minnesota,
Minneapolis, MN 55455, USA}\\[1mm]
$^4${\it  William I. Fine Theoretical Physics Institute,
University of Minnesota,
Minneapolis, MN 55455, USA}\\[1mm]
$^{5}${\it Kavli Institute for Theoretical Physics, University of California,Santa Barbara, CA 93106, USA}

\end{center}

\vspace{0.6cm}

\begin{center}
{\large\bf Abstract}
\end{center}
Assuming the Lorentz and {\bf CPT} invariances we show that neutron-antineutron oscillation 
implies  breaking of {\bf CP} along with baryon number violation -- 
i.e. two of Sakharov conditions for baryogenesis.
The oscillation is produced by the unique operator in the effective Hamiltonian.
This operator mixing neutron and antineutron preserves charge conjugation {\bf C}
and breaks {\bf P} and {\bf T}.  External magnetic field always leads to suppression of oscillations.
Its presence does not lead to any new operator mixing neutron and antineutron.\\[10cm]
\end{titlepage}

\newpage

\title{Neutron--Antineutron Oscillation as a Signal of CP Violation}

\author{Zurab Berezhiani} 
\affiliation{Dipartimento di Fisica, Universit\`a dell'Aquila, Via Vetoio, 67100 Coppito, L'Aquila, Italy} 
\affiliation{INFN, Laboratori Nazionali Gran Sasso, 67010 Assergi,  L'Aquila, Italy}

\author{Arkady Vainshtein}
\affiliation{Department of Physics, University of Minnesota,
Minneapolis, MN 55455, USA}
\affiliation{William I. Fine Theoretical Physics Institute,
University of Minnesota,
Minneapolis, MN 55455, USA}
\affiliation{Kavli Institute for Theoretical Physics, University of California,Santa Barbara, CA 93106, USA}


\begin{abstract}
Assuming the Lorentz and {\bf CPT} invariances we show that neutron-antineutron oscillation 
implies  breaking of {\bf CP} along with baryon number violation -- 
i.e. two of Sakharov conditions for baryogenesis.
The oscillation is produced by the unique operator in the effective Hamiltonian.
This operator mixing neutron and antineutron preserves charge conjugation {\bf C}
and breaks {\bf P} and {\bf T}.  External magnetic field always leads to suppression of oscillations.
Its presence does not lead to any new operator mixing neutron and antineutron.
 \end{abstract}

\maketitle

\noindent
{\bf1.} Experimental search for neutron-antineutron oscillation \cite{Kuzmin:1970nx}
 is under active discussion nowadays
(see the resent review\,\cite{Phillips:2014fgb}). Its discovery  
would be a clear evidence of baryon charge nonconservation, $|\Delta {\cal B} |=2$.
In this note we would like to emphasize that neutron-antineutron  oscillation also
breaks {\bf CP} invariance. This conclusion is based on the Lorentz invariance and {\bf CPT}.
  
To demonstrate our assertion let us start with the Dirac Lagrangian
\begin{equation}
{\cal L}=i\bar n \gamma^{\mu} \partial_{\mu}n - m\,\bar n n
\label{lagr}
\end{equation}
with four-component spinor $n$ and the mass parameter $m$ which is real and positive.  
The Lagrangian
 gives the Lorentz-invariant description
of free neutron and antineutron states and preserves
the baryon charge, ${\cal B}=1$ for $n$ and ${\cal B}=-1$
for $\bar n$. This charge corresponds to the continuous symmetry
\begin{equation}
n \to {\rm e}^{i\alpha }n , \quad \bar n \to {\rm e}^{-i\alpha }\bar n 
\label{U1B}
\end{equation}
of Lagrangian (\ref{lagr}). At each spatial momentum there are four degenerate states,
two spin doublets which differ by the baryon charge ${\cal B}$. 

Another bilinear mass term,
\begin{equation}
\Delta {\cal L}_{m^{\prime}}=-im' \bar n \gamma_{5} n\,,
\label{deltam}
\end{equation}
consistent with the baryon
charge conservation, can be rotated away by chiral transformation $n\to  {\rm e}^{i\beta \gamma_{5}}n$
if there is no terms breaking the baryon charge. As we will see it is not the case when the baryon
charge is broken.  

How the baryon number non-conservation shows up at the level of free one-particle 
states? In Lagrangian description it could be only modification of the bilinear mass term.
We show below that the most generic Lorentz invariant modification of Eq.\,(\ref{lagr}) reduces to 
one possibility for the baryon charge breaking  by two units,
\begin{equation}
\Delta {\cal L}_{{\cal B}\!\!\!\!\not}~=-\frac 1 2 \,\epsilon \,\big [n^{T}\!Cn +\bar n C {\bar n}^{T}\,\big ]\,.
\label{deltaB}
\end{equation}
Here $C=i\gamma^{2}\gamma^{0}$ is the charge conjugation matrix 
in the standard representation of gamma matrices, and 
$\epsilon$ is a real positive parameter. The reality of $\epsilon$ 
as a coefficient for $n^{T}Cn$ can be always achieved by the phase rotation
(\ref{U1B}) of $n$ field.

One could add also $|\Delta {\cal B} |=2$ term of the form $n^{T}C\gamma_{5} n$. 
However, it can be rotated away by the chiral rotation $n\to  {\rm e}^{i\beta \gamma_{5}}n$.
The price for this is, as we mentioned above, an appearance of the $\gamma_{5}$
mass term (\ref{deltam}).  Also mixed kinetic terms 
$\propto  i \bar n \gamma^{\mu} C \partial_{\mu} \bar{n}^T + {\rm h.c.}$ can be turned away 
with redefinition of the fermion field. 

Hence,  a generic Lagrangian containing the fermion bilinears can always be brought  
to a form containing only the terms (\ref{lagr}), (\ref{deltam}) and (\ref{deltaB}). 

What is the status of discrete {\bf C}, {\bf P} and {\bf T} symmetries 
 in this situation? 
 It is simple to verify that the Lagrangian terms (\ref{lagr}), (\ref{deltam}) and (\ref{deltaB}) 
are all invariant under the charge conjugation {\bf C}, 
\begin{equation}
n \to n^{c}=C{\bar n}^{T}\,.
\label{conj}
\end{equation}
In fact, the expression (\ref{deltaB})  can be rewritten in the form 
$
-(1/2) \,\epsilon \,\big [\xbar{n^{c}}\, n +\bar n \, n^{c}\,\big ],
$
which makes its ${\bf C}$ invariance explicit.

The parity transformation {\bf P} involves (besides reflection of the space coordinates)
the substitution 
\begin{equation}
n \to \gamma^{0}n\,.
\label{parity}
\end{equation}
This substitution changes $\Delta {\cal L}_{{\cal B}\!\!\!\!\not}~$ to $-\Delta {\cal L}_{{\cal B}\!\!\!\!\not}~$ because $\gamma^{0}C\gamma^{0}=-C$.
The breaking of parity in neutron-antineutron transition reflects the well-known feature of the opposite parity of fermion
and antifermion. The term $\Delta {\cal L}_{m^{\prime}}$ also breaks {\bf P} parity, it is evidently pseudoscalar.
Clearly, the parity violation 
comes together with breaking of {\bf T} invariance since {\bf CPT} invariance 
is guaranteed by a local, Lorentz invariant form of the Lagrangian. 

Thus, we demonstrated that observation of neutron-antineutron oscillation signals breaking of {\bf CP} invariance together
with breaking of baryon charge. \\[1mm]

\noindent
{\bf 2.} To show that the above consideration covers indeed a generic case it is convenient
to introduce two left-handed Weyl spinors \cite{Ramond:1999vh}, forming a flavor doublet 
\begin{equation}
\psi^{i\,\alpha}\,,\qquad i=1,2,\quad \alpha=1,2\,,
\label{doubl}
\end{equation}
together with their complex conjugates, representing the right-handed spinors,
\begin{equation}
\xbar{\psi}^{\,\dot\alpha}_i= (\psi^{i\,\alpha})^\ast \,,\qquad i=1,2,\quad \dot\alpha=1,2\,.
\end{equation}
One can raise and lower space $\alpha, \dot\alpha$ and flavor $i$ indices using $\epsilon_{\alpha\beta},\epsilon_{\dot\alpha\dot\beta}$
and $\epsilon_{ik}$.
In terms of Dirac spinor $n$ these two left-handed Weyl spinors are $n_{L}$ and $(n_{R})^{*}$.
The most generic Lagrangian is
\begin{equation}
{\cal L}=\psi^{i\,\alpha} \,i\partial_{\alpha\dot\alpha} \xbar{\psi}^{\,\dot\alpha}_i\,
+ \frac 12\, \Big[m_{ik}\psi^{i\alpha}\psi^{k}_\alpha+\xbar{m}^{ki}\xbar{\psi}_{k\,\dot\alpha}
\xbar{\psi}^{\,\dot\alpha}_i\Big]\,,
\label{weyl}
\end{equation}
where $\partial_{\alpha\dot\alpha}=\sigma^{\mu}_{\alpha\dot\alpha}\partial_{\mu}$, $\sigma^{\mu}=\{1,\vec \sigma\}$,  $m_{ik}$ is the symmetric mass matrix, $m_{ik}=m_{ki}$ and 
$\xbar{m}^{ik}=(m_{ik})^{*}$ is its conjugate.

The kinetic term in (\ref{weyl}) is U(2) symmetric: besides SU(2) rotations of the doublets,
there is U(1) associated with the overall phase rotation of the doublet (\ref{doubl}).
The mass terms break both, U(1) and SU(2) flavor symmetries, so, generically, no 
continuous symmetry remains. 

To see how the symmetry (\ref{U1B}) associated with the baryon charge
could arise note that one can interpret U(2) transformations as acting on the external 
mass matrix $m_{ik}$. This matrix is charged under U(1), the overall phase rotation, so
this U(1) symmetry is always broken. In respect to SU(2) transformations the symmetric
tensor $m_{ik}$ is the adjoint representation, i.e., can be viewed as an isovector $\mu^{a}$, $a=1,2,3$,
\begin{equation}
m^i_k=\varepsilon^{ij} m_{jk} =  \mu^a (\tau^a)^i_k\,,\quad a=1,2,3\,, 
\label{defmu}
\end{equation}
Because $\mu^{a}$ is complex, we are actually dealing with two real isovectors,
 ${\rm Re}\,\mu^{a}$ and  ${\rm Im}\,\mu^{a}$. 
The SU(2) transformations are equivalent to simultaneous rotation of real and imaginary 
vectors, while U(1) changes phases of all $\mu^{a}$ simultaneously, which is equivalent  
to SO(2)  rotation inside each couple ${\rm Re}\,\mu^{a}$, ${\rm Im}\,\mu^{a}$. 
 Only in case when these 
 vectors are parallel we have an invariance of the mass matrix which is just
 a rotation around this common direction. 
 (In this case, all ${\rm Im}\,\mu^{a}$ can be absorbed in ${\rm Re}\,\mu^{a}$ by 
 U(1) transformation.) 
 This symmetry is the one
 identified with the baryonic U(1) in Eq.\,(\ref{U1B}).

Let us show now that in the absence of the common direction we get 
two spin 1/2 Majorana fermions with different masses.
From equations of motion
\begin{equation}
\begin{split}
i\,\partial_{\alpha\dot\alpha} \psi^{i\,\alpha}\!+\!
\xbar{m}^{ik}\xbar{\psi}_{k\,\dot\alpha}\!=0\,,\\
i\,\partial_{\alpha\dot\alpha}\xbar{\psi}^{\,\dot\alpha}_i-m_{ik}\,\psi^k_\alpha=0~
\end{split}
\end{equation}
to exclude $\xbar{\psi}^{\,\dot\alpha}_i$ we come to the eigenvalue problem for
$M^2=p_\mu p^\mu$,
\begin{equation}
M^2\psi^{k\alpha}-\xbar{m}^{kl}m_{ln}\psi^{n\alpha}=0\,.
\end{equation}

Using definition (\ref{defmu}) of $\mu^{a}$ the squared mass matrix 
can be presented as a combination of isoscalar and isovector pieces:  
\begin{equation}
\xbar{m}^{kl}m_{ln}=\mu^a\xbar{\mu}^{\,a}\,\delta^k_n  +i\epsilon^{abc}\mu^a\xbar{\mu}^{\,b} (\tau^c)^k_n\,.
\label{mu2}
\end{equation}
Correspondingly, there are two invariants defining $M^2$. The isoscalar part gives the sum of eigenvalues,
\begin{equation}
\frac{M_1^2+M_2^2}{2}=\mu^a\xbar{\mu}^{\,a}= ({\rm Re} \,\mu^a)^2+ ({\rm \,Im}\,\mu^a)^2\, 
\end{equation}
while the length of the isovector part defines the splitting of the eigenvalues,
\begin{equation}
\frac{M_1^2-M_2^2}{2}=2\sqrt{\Big[\epsilon^{abc} \,{\rm Re} \,\mu^a {\rm \,Im}\,\mu^b\Big]^2}\,.
\end{equation}
Thus, we see the splitting associated with the breaking of the baryon charge.

To follow the discrete symmetries we can orient the  mass matrix $m_{ik}$ in a convenient way.
In terms of $\mu^{a}$ the matrix has the form
\begin{equation} 
m_{ik} = \mat{-\mu^{1}- i\mu^{2} }{\mu^{3} }{\mu^{3}}{\mu^{1} - i\mu^{2}}  . 
\label{mass-matrix}
\end{equation} 
Without lost of generality we can put both, ${\rm Re}\,\mu^{a}$ and  ${\rm Im}\,\mu^{a}$, onto the 23 plane,
i.e., put $\mu^{1}=0$. Moreover, we can orient ${\rm Re}\,\mu^{a}$ along the the third axis, i.e., put ${\rm Re}\,\mu^{2}=0$.
Then, only 3 nonvanishing parameters, ${\rm Re}\,\mu^{3}$, ${\rm Im}\,\mu^{3}$ and ${\rm Im}\,\mu^{2}$, remain
and the mass matrix takes the form,
\begin{equation} 
m_{ik} = \mat{{\rm Im}\,\mu^{2} }{{\rm Re}\,\mu^{3}+i\,{\rm Im}\,\mu^{3} }{{\rm Re}\,\mu^{3}+i\,{\rm Im}\,\mu^{3}}{{\rm Im}\,\mu^{2}}  . 
\label{mass-matrix1}
\end{equation} 

In the Weyl description the charge conjugation {\bf C} 
is just an interchange of $\psi^{1\,\alpha}$ and $\psi^{2\,\alpha}$, the symmetry 
which implies that $m_{12}=m_{21}$ and $m_{11}=m_{22}$.
The matrix (\ref{mass-matrix1}) clearly satisfies these conditions.
 The {\bf P} reflection involves the interchange
$
\psi^{i\alpha} \leftrightarrow \bar\psi^{\,i\dot\alpha}=\epsilon^{ik}\bar\psi_k^{\,\dot\alpha}\,.
$
This symmetry is broken by nonvanishing ${\rm \,Im}\,\mu^a$.

Now it is simple to establish a correspondence with parameters introduced 
earlier in four-component spinor notations. Namely,
\begin{equation} 
{\rm Re}\,\mu^{3}=m\,,\quad {\rm Im}\,\mu^{3}=m^{\prime}\,,\quad {\rm Im}\,\mu^{2}=\epsilon\,.
\label{paramcor}
\end{equation} 
So while {\bf C} parity is preserved, we have {\bf P} even, Eq.\,(\ref{lagr}), and {\bf P} odd,  Eqs.\,(\ref{deltam}), (\ref{deltaB}), mass terms. 
Thus, we proved for generic case the association of baryon charge breaking with {\bf CP} violation.

Note that in terms of remaining 3 parameters the masses of {\bf C} even and {\bf C} odd
Majorana fermions are
\begin{equation}
M_{1}^{2}=(m+\epsilon)^{2}+(m^{\prime})^{2}\,,\quad M_{2}^{2}=(m-\epsilon)^{2}+(m^{\prime})^{2}\,,
\end{equation}
what different from standard expressions when $m^{\prime}$ is nonvanishing.
In particular, it implies that the oscillation time $\tau_{n\bar n}$ in free neutron transition probability,
$P_{n\bar n}(t) = \sin^2(t/\tau_{n\bar n}) $ is $\sqrt{1+(m^{\prime}/m)^{2}}/\epsilon$ instead 
of $1/\epsilon$.

The {\bf CP} odd nature of the operator (\ref{deltaB}) was noted recently in Ref.~\cite{Gardner:2014cma}.
However, the authors of this paper discussed also the {\bf CP} even operator $n^{T}\gamma_{5}Cn$ which,
as we showed, can be rotated away by field redefinition. These authors also analyzed modifications induced by 
external magnetic field claiming an existence of a new $n-\bar n$ transition magnetic moment and also an absence 
of the usual suppression of $n-\bar n$ oscillation in presence of magnetic field. We will show below that both claims
are invalid.
\\[1mm]

\noindent 
{\bf 3.} Our consideration above refers to the neutron-antineutron oscillation in vacuum.
Now we show that even in the presence of magnetic field no new $|\Delta {\cal B}|=2$ operator
appears. Similar consideration  was done  in Ref.~\cite{Voloshin:1987qy}   in application to 
magnetic moment of neutrinos.

In the Weyl formalism the field strengths tensor $F_{\mu\nu}$ is substituted by the symmetric 
tensor $F_{\alpha\beta}$ and its complex conjugate $\bar F_{\dot\alpha\dot\beta}$. They correspond
to $\vec E\pm i \vec B$ combinations of electric and magnetic fields. Then Lorentz invariance allows only
two structures involving electromagnetic fields,
\begin{equation}
F_{\alpha\beta}\psi^{i\alpha}\psi^{k\beta}\epsilon_{ik}\,,\quad \bar F_{\dot\alpha\dot\beta}\bar\psi_{i}^{\dot\alpha}\bar\psi_{k}^{\dot\beta}\epsilon^{ik}\
\end{equation}
Antisymmetry in flavor indices implies that spinors with the opposite baryon charge enter. So both operators preserve the baryon 
charge, they describe interactions with the magnetic and electric dipole moments of the neutron. 

The authors of \cite{Gardner:2014cma} realize that the operator $n^{T}\sigma^{\mu\nu} CnF_{\mu\nu}$ is vanishing 
due to Fermi statistics. They believe, however, that a composite nature of neutron changes the situation and a new type of magnetic moment in $\Delta {\cal B}=\pm 2$ transitions may present. In other words they think that the effective Lagrangian description is broken for composite particles. 

To show that is not the case let us consider the process
\begin{equation}
n(p_{1})+n(p_{2}) \to \gamma^{*}(k)
\label{nngamma}
\end{equation}
in the crossing channel to $n-\bar n \gamma^{*}$ transition.  The number of invariant amplitudes for
the process (\ref{nngamma}) which is $1/2^{+}+1/2^{+}\to 1^{-}$ transition is equal to one.
Only orbital momentum $L=1$ and total spin $S=1$ in two neutron system are allowed by angular 
momentum conservation and Fermi statistics. The gauge-invariant form of the  amplitude is
\begin{equation}
u^{T}\!(p_{1})C\gamma^{\mu}\gamma_{5}u(p_{2}) \,k^{\nu}F_{\mu\nu}\,,\qquad F_{\mu\nu}=k_{\mu}\epsilon_{\nu}-k_{\nu}\epsilon_{\mu},
\label{nngamma1}
\end{equation}
where $u_{1,2}$ are Dirac spinors describing neutrons and $\epsilon_{\mu}$ refers to the gauge potential.
In space representation we deal with $\partial^{\nu}F_{\mu\nu}$ which vanishes outside of the source of 
the electromagnetic field, and, in particular, for the distributed magnetic field. It proves that there is no place for magnetic moment of $n-\bar n$ transition, and effective 
Lagrangian description does work.\footnote{Let us also remark that $n-\bar n \gamma^{*}$ transition 
with a virtual photon connected to the proton, as well as $nn \to\gamma^{*}$ 
annihilation, would destabilise the nuclei even in the absence of $n-\bar n$ mass mixing.}


Even in the absence of new $n-\bar n$ magnetic moment the authors of \cite{Gardner:2014cma} 
claim that suppression of $n-\bar n$ oscillations by external magnetic field can be overcome
by applying the magnetic field transversal to quantization axis. 

In their first example where the transversal field is time -independent (after switching) they
obtained four different energy eigenvalues (Eq.\,(26) in \cite{Gardner:2014cma}) which depend on direction
of magnetic field. This clearly breaks rotational invariance. The source of this breaking is the wrong sign
of the ${\cal H}_{34}$ and ${\cal H}_{43}$ in the Hamiltonian matrix ${\cal H}$ in Eq.\,(20) 
The existing sign implies 
that $\Delta {\cal B}=2$ amplitude is of different sign for spins up and down. Changing sign of ${\cal H}_{34}$ and ${\cal H}_{43}$ restores rotational invariance, the eigenvalues become 
$E=M_{1}\pm\sqrt{\omega_{0}^{2}+\omega_{1}^{2}+\delta^{2}}$, each doubly degenerate.
In their second example, where the transversal field is rotating, the result of \cite{Gardner:2014cma} is also incorrect --
after a change of variables indicated in \cite{Gardner:2014cma} the consideration is similar to the first example with 
time-independent field.

As a consequence the magnetic field suppression does present indeed, and the suggestion in \cite{Gardner:2014cma} that $n-\bar n$ oscillations can be measured without minimizing 
magnetic field does not work.\footnote{ 
The situation is different if  one considers oscillation $n-n'$ where $n'$ is a mirror 
neutron,  twin of the neutron from hidden mirror sector  \cite{Berezhiani:2005hv}. 
In this case, operators  $\overline{n}\sigma^{\mu\nu} n' F_{\mu\nu}$ and/or  
 $n^{T}\sigma^{\mu\nu} Cn' F_{\mu\nu}$ are allowed. Hence,  $n-n'$ and/or $n-\bar n'$ 
 transition probabilities may not depend on the value of magnetic field provided that it is large enough, 
 with possible implications for the experimental search of neutron$-$mirror neutron oscillations.}\\[1mm]

\noindent
{\bf 4.} Our use of the effective Lagrangian for the proof means that the Lorentz invariance and {\bf CPT}
are crucial inputs. Once constraints of Lorentz invariance are lifted new $|\Delta {\cal B}|=2$ operators could show up. 

Such operators were analyzed in Ref.\,\cite{Babu:2015axa} for putting limits on the Lorentz invariance breaking.  
In particular, the authors suggested the operator $n^{T}C\gamma^{5}\gamma^{2}n$ as an example
which involves spin flip and, correspondingly, less dependent on magnetic field surrounding.

Note, however, that besides breaking of Lorentz invariance this operator breaks also 3d rotational invariance, i.e.,
isotropy of space. Such anisotropy could be studied by measuring spin effects in neutron-antineutron transitions.\\[1mm]

\noindent
{\bf 5.} 
In the Standard Model (SM) conservations of baryon ${\cal B}$ and lepton ${\cal L}$ numbers  
are related to accidental global symmetries of the SM Lagrangian. (Nonperturbative
breaking of ${\cal B}$ and ${\cal L}$, preserving ${\cal B}-{\cal L}$, is extremely small.)
The violation of ${\cal B}$ by two units can be originated only from new physics beyond SM which 
could induce the effective six-quark operators  
\begin{equation}
{\cal O}= \frac{1}{M^5}\, udd udd 
\label{D9}
\end{equation}
involving $u$ and $d$ quarks of different families 
in different color and Lorentz invariant combinations 
(all possible convolutions of spinor indices are omitted). The smallness of baryon violation 
is related to the large mass scale $M$ related to new physics.

\begin{figure}[t]
\vspace{-3cm}
\begin{center}
\includegraphics[width=7cm]{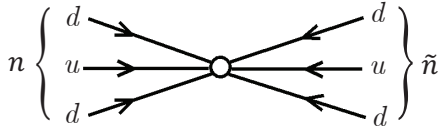}
\vskip -4cm
\caption{
\label{fig1}
Diagram for generating  $n - \bar n$ mixing terms }
\end{center}
\end{figure}

In fact, the ${\cal B}$ breaking mass term (\ref{deltaB})
emerges  by taking matrix element between $n$ and $\bar n$ states 
of the operator structures (\ref{D9}), see diagram in Fig.\,\ref{fig1},
\begin{equation} 
-\frac 1 2 \,\epsilon \,\langle \bar n | n^{T}\!Cn  |n\rangle =
\langle \bar n | {\cal O} |n\rangle \,.
\end{equation}
It gives an estimate of order $\Lambda_{\rm QCD}^6/M^5$ for the parameter 
$\epsilon$ which describes the 
oscillation time. 
\begin{figure}[t]
\vspace{-0.7cm}
\begin{center}
\includegraphics[width=6cm]{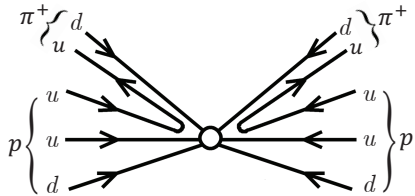}
\vskip -3.5cm
\caption{
\label{fig2}
Inducing  $pp \to \pi^+ \pi^+$ annihilation via operators (\ref{D9}) }
\end{center}
\end{figure}

Our consideration shows that only operators which are {\bf C} even and {\bf P} odd
contribute to the above matrix element (up to small corrections due to electroweak 
interactions where the discrete symmetries are broken). 
In general, operators coming from physics beyond SM
do not respect any of discrete symmetries {\bf C}, {\bf P} and {\bf CP}.  
If, however, a new physics model  produces ${\cal B}$ violating operators which do not satisfy
the selection rules of $n-\bar n$ transition, their effect will show up in instability of nuclei but not 
in free neutron-antineutron oscillations. Indeed, such operators would induce  processes 
{of annihilation of two nucleons }
like $N+N \to \pi+\pi$ inside nucleus,  as shown on Fig.\,\ref{fig2}.

The operators of the type of (\ref{D9}) involving strange quark, $udsuds$, could induce $\Lambda-\bar\Lambda$  mixing. However, such operators  would also lead  to nuclear instability via nucleon annihilation into kaons $N + N \to K + K$, see the diagram in Fig.\,\ref{fig2} where 
in upper lines $d$ quark is substituted by $s$ quark (and $\pi^{+}$ by $K^{+}$). 
In fact, nuclear instability bounds on 
$\Lambda-\bar\Lambda$ mixing are only mildly, within an order of magnitude, weaker 
than with respect to $n-\bar n$ mixing which makes hopeless the possibility to detect 
 $\Lambda-\bar\Lambda$ oscillation in the hyperon beam. 
(Instead, it can be of interest to search for the nuclear 
decays into kaons  in the large volume detectors.)
The nuclear instability limits on $\Lambda-\bar\Lambda$ mixing are 
about 15 orders of magnitude stronger than the sensitivity 
$\delta_{\Lambda\bar\Lambda}\sim 10^{-6}$~eV which can be achieved in the 
laboratory conditions \cite{Kang}. The nuclear stability limits make hopeless also 
the laboratory search of $bus$-like baryon oscillation due to operator 
$usbusb$ suggested in Ref. \cite{Kuzmin}.  
\\[1mm]

\noindent
{\bf 6.}
The construction we used for neutron-antineutron transition could be applied to mixing of
massive neutrinos. As an example, let us take the system of left-handed $\nu_{e}$ and
$\nu_{\mu}$ and their conjugated partners, right-handed $\bar\nu_{e}$ and
$\bar\nu_{\mu}$. One can ascribe them \cite{Konopinski:1953gq} a flavor charge ${\cal F}={\cal L}_{e}-{\cal L}_{\mu}$ (analog of ${\cal B}$), to be (+1) for $\nu_{e}$  and (-1)
for $\nu_{\mu}$. Then, {\bf C} conjugation is interchange of $\nu_{e}$ and
$\nu_{\mu}$. Again, ${\cal F}$ breaking mass term would be {\bf C} even and {\bf P} odd.

A similar scenario can be played in case of Dirac massive neutrino. \\[1mm]

\noindent
{\bf 7.} In summary, we show that the Lorentz and {\bf CPT} invariance lead to the unique 
$|\Delta {\cal B}|=2$ operator in
the neutron-antineutron mixing. This operator is {\bf CP} odd. Switching on external magnetic field  influences the level splitting
what suppresses $n-\bar n$ oscillations but does not add any new $|\Delta {\cal B}|=2$ operator in contradistinction 
with recent claims in literature.

Interesting to note that observation of neutron-antineutron transition would show that two of three 
Sakharov conditions for baryogenesis are satisfied,  violations of ${\cal B-L}$ and {\bf CP}.  
However, it would be honest to say that primordial baryogenesis in the Early Universe 
should be related to underlying physics that induces operators (\ref{D9}) rather than 
to  neutron-antineutron oscillation phenomenon itself. 
On the other hand, for new physics involving contact operators (\ref{D9}) (or heavy particles
mediating these operators) the third, out-of-equilibrium condition is also automatically
satisfied when the universe temperature drops below the relevant mass scales. 
Thus, discovery of neutron-antineutron oscillation would make it manifest that these operators 
contain {\bf CP} violating terms which could be 
at the origin of the baryon asymmetry of the Universe. \\[6mm]

We thank Susan Gardner, Yuri Kamyshkov, Kirill Melnikov, Rabi Mohapatra and Misha Voloshin for helpful discussions.
A.V. appreciates hospitality of the Kavli Institute for Theoretical Physics where his research was supported 
in part by the National Science Foundation under Grant No.\ NSF PHY11-25915. 
The work of Z.B. was supported in part by the MIUR 
triennal grant for Research Projects of National Interest  PRIN No. 2012CPPYP7 
``Astroparticle Physics", and in part by 
Rustaveli National Science Foundation grant No. DI/8/6-100/12.



\vskip 1.5cm




\begin{thebibliography}{99}

\bibitem{Kuzmin:1970nx} 
  V.\,A.~Kuzmin,
  Pisma Zh.\ Eksp.\ Teor.\ Fiz.\  {\bf 12}, 335 (1970); 
  R.~N.~Mohapatra and R.~E.~Marshak,
  Phys.\ Rev.\ Lett.\  {\bf 44}, 1316 (1980).


\bibitem{Phillips:2014fgb} 
  D.\,G.~Phillips, II, W.\,M.~Snow, K.~Babu, 
  {\it et al.},
  ``Neutron-Antineutron Oscillations: Theoretical Status and Experimental Prospects,''
  [arXiv:1410.1100 [hep-ex]]

  
\bibitem{Ramond:1999vh} 
See e.g. in  P.~Ramond,  ``Journeys Beyond the Standard Model,''
  Reading, Mass., Perseus Books, 1999, where the Weyl formalism 
  is gracefully  applied to description of massive neutrinos.

\bibitem{Gardner:2014cma} 
  S.~Gardner and E.~Jafari,
  Phys.\ Rev.\ D {\bf 91}, no. 9, 096010 (2015)
  [arXiv:1408.2264 [hep-ph]].
  
\bibitem{Voloshin:1987qy} 
 M.\,B.~Voloshin,
  Sov.\ J.\ Nucl.\ Phys.\  {\bf 48}, 512 (1988)
 [Yad.\ Fiz.\  {\bf 48}, 804 (1988)].
  
\bibitem{Berezhiani:2005hv} 
  Z.~Berezhiani and L.~Bento,
  Phys.\ Rev.\ Lett.\  {\bf 96}, 081801 (2006)
  [hep-ph/0507031]; 
  Z.~Berezhiani,
  Eur.\ Phys.\ J.\ C {\bf 64}, 421 (2009)
  [arXiv:0804.2088 [hep-ph]].
    
\bibitem{Babu:2015axa} 
  K.\,S.~Babu and R.\,N.~Mohapatra,
  Phys.\ Rev.\ D {\bf 91}, no. 9, 096009 (2015)
  [arXiv:1504.01176 [hep-ph]].

\bibitem{Kang} 
  X.\,W.~Kang, H.\,B.~Li and G.\,R.~Lu,
  Phys.\ Rev.\ D {\bf 81}, 051901 (2010)
  [arXiv:0906.0230 [hep-ph]].
  
\bibitem{Kuzmin} 
  V.\,A.~Kuzmin,
  In *Oak Ridge 1996, Future prospects of baryon instability search* 89-91
  [hep-ph/9609253].  

\bibitem{Konopinski:1953gq} 
Ya.\,B.~Zeldovich, Dokl. Akad. Nauk SSSR {\bf 86}, 505 (1952); 
  E.\,J.~Konopinski and H.\,M.~Mahmoud,
  Phys.\ Rev.\  {\bf 92}, 1045 (1953).
 

    
\end{thebibliography}
 \end{document}